\documentclass[fleqn,usenatbib]{mnras}

\usepackage[usenames,dvipsnames]{xcolor}
\usepackage{amsmath}
\usepackage{amssymb}
\usepackage{array}
\usepackage{xfrac}
\usepackage{multirow}

\newcommand{\rev}[1]{{\textcolor{black}{#1}}}
\usepackage[T1]{fontenc}
\usepackage{ae,aecompl}
\hypersetup{draft}

\title[A tight N/O--potential relation]{A tight N/O--potential relation in star-forming galaxies} 

\author[Boardman et al.]{
N.~Boardman$^{1}$\thanks{E-mail: nfb@st-andrews.ac.uk},
V.~Wild$^{1}$,
N.~Vale Asari$^{2}$\\
$^{1}$School of Physics and Astronomy, University of St Andrews, North Haugh, St Andrews KY16 9SS, UK\\
$^{2}$Departamento de F\'{\i}sica--CFM, Universidade Federal de Santa Catarina, C.P.\ 5064, 88035-972, Florian\'opolis, SC, Brazil\\
}

\date{Accepted X. Received X; in original form X}
\pubyear{2022}
\begin{document} 
\label{firstpage}
\pagerange{\pageref{firstpage}--\pageref{lastpage}}
\maketitle

\begin{abstract}

We report a significantly tighter trend between gaseous N/O and $M_*/R_e$ (a proxy for gravitational potential) than has previously been reported between gaseous metallicity and $M_*/R_e$, for star-forming galaxies in the MaNGA survey. We argue this result to be a consequence of deeper potential wells conferring greater resistance to \rev{metal outflows while also being associated with earlier star-formation histories}, combined with N/O being comparatively \rev{unaffected by} metal-poor inflows. The potential--N/O relation thus appears \rev{to be both} more resistant to short-timescale baryonic processes \rev{and also more reflective of a galaxy's chemical evolution state, when compared to} previously-considered relations. 

\end{abstract}


\section{Introduction}\label{intro}

The question of chemical evolution remains at the heart of observational galaxy astrophysics. The chemistry of galaxies' stars and gas have been studied in much detail, in terms of both metallicities and in terms of elemental abundance ratios. For stars, the most frequently considered abundance measures are $\mathrm{[Fe/H]}$ and $\mathrm{[\alpha/Fe]}$; for gas, the most common measures are instead $\mathrm{12 + \log(O/H)}$ and $\mathrm{\log(N/O)}$. In both cases, combining measures yields much more information than does any one measure alone, due to the different enrichment timescales involved. Alpha elements such as oxygen are released into the ISM via Type-II supernovae, resulting in enrichment on just $\sim$10 Myr timescales after a star-formation event \citep[e.g.][]{timmes1995}; iron and nitrogen meanwhile are released on much longer timescales, largely by Type-Ia supernovae and by AGB stars respectively \citep[see][for a review]{maiolino2019}. N/O and metallicity have been shown repeatedly to be tightly correlated at higher metallicities, with N/O instead displaying a near-constant value in low-metallicity objects \citep[e.g.][]{pilyugin2010,andrews2013,pm2016}. 

Scaling relations provide important constraints for chemical evolution models and thus provide important insight into the relevant physics. Of particular importance is the mass--metallicity relation, in which metallicity rises with increasing stellar mass ($M_*$), which has been noted repeatedly for both stellar and gaseous metallicities \citep[e.g.][]{lequeux1979,tremonti2004,gallazzi2005,kewley2008}. The gaseous mass--metallicity relation flattens at high masses, potentially indicating such galaxies to have reached a metallicity equilibrium \citep[e.g.][]{lilly2013,belfiore2019a}. More compact galaxies have also been reported to be more metal-rich at a given stellar mass in both observations \citep[e.g.][]{ellison2008,mcdermid2015,li2018} and simulations \citep[][]{ma2024}, with galaxies' size typically parameterised using the half-light radius $R_e$, and more compact galaxies also appear to possess flatter metallicity gradients over a wide range of stellar masses \citep{boardman2021,boardman2022}. The parameter $M_*/R_e$ (hereafter $\mathit{\Phi_e}$) has been reported to correlate particularly tightly with galaxies' stellar or gaseous metallicities, with $M_*$ and $M_*/R_e^2$ (hereafter $\mathit{\Sigma_e}$) producing looser correlations by comparison \citep{deugenio2018,barone2018,barone2020,barone2022,vaughan2022,ma2024,sm2024}. This has been repeatedly interpreted as reflecting the importance of gravitational potential, for which $\mathit{\Phi_e}$ is considered a proxy; in this scenario, higher gravitational potentials confer greater resistance to metal-loss via outflows, resulting in great chemical enrichment in more compact systems. \citet{baker2023a} have however challenged this interpretation, reporting a comparatively weak metallicity trend when dynamical mass is considered instead of stellar mass. 

In this letter we report a particularly tight correlation between $\mathrm{\log(N/O)}$ and $\mathit{\Phi_e}$, using data from the SDSS-IV MaNGA integral-field unit (IFU) survey. We further report that $\mathit{\Phi_e}$ correlates more tightly with $\mathrm{\log(N/O)}$ than with gaseous metallicity, with the $\mathit{\Phi_e}$--metallicity correlation largely disappearing once  $\mathrm{\log(N/O)}$ is accounted for.  We present our sample and data in \autoref{sampledata}, and our results in \autoref{results}. We discuss our findings and conclude in \autoref{disc}. We use `metallicity' to refer to gas-phase metallicities $\mathrm{12 + \log(O/H)}$, assume a \citet{chabrier2003} initial mass function and adopt the following $\Lambda$ Cold Dark Matter cosmology: $\mathrm{H_0} = 71$ km/s/Mpc, $\mathrm{\Omega_M} = 0.27$, $\mathrm{\Omega_\Lambda} = 0.73$.

\section{Sample \& data}\label{sampledata}

\begin{figure}
\begin{center}
	\includegraphics[trim = 0.7cm 10.9cm 1cm 9.8cm,scale=0.7,clip]{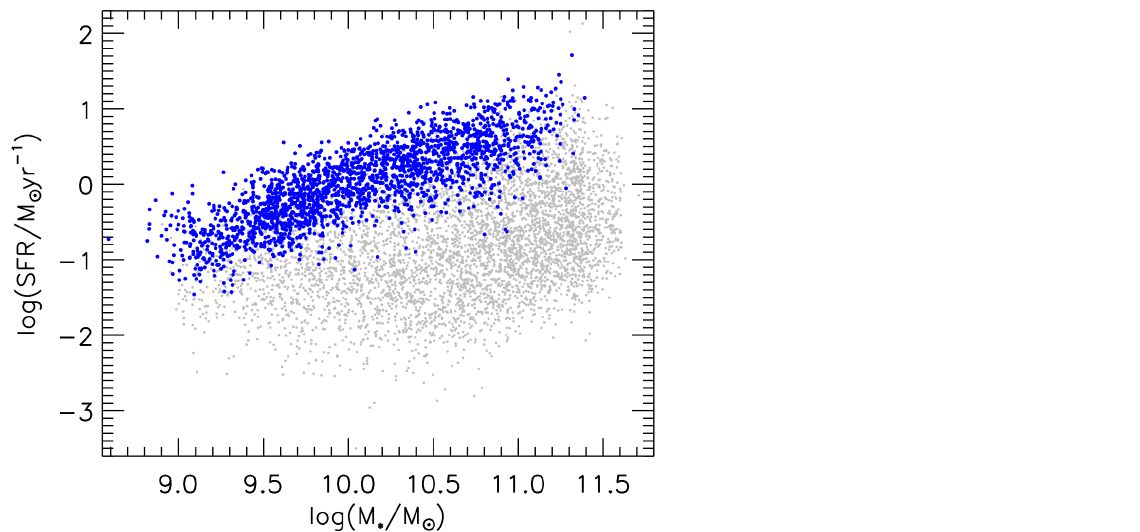} 
	\caption{Selected sample (blue points) and MaNGA parent sample (grey points) in terms of stellar mass ($h=0.71$) and star-formation rate (Chabrier IMF).}
	\label{sfms}
	\end{center}
\end{figure}

We draw our parent galaxy sample from the SDSS-IV MaNGA survey \citep{bundy2015}, for which all data is publically available as of SDSS Data Release 17 \citep[DR17][]{sdssdr17}. The MaNGA survey performed IFU spectroscopic observations on $\sim$10000 galaxies out to redshifts $z \lesssim 0.15$ \citep{yan2016b,wake2017}, with the main MaNGA sample selected to possess a roughly flat distribution in $\log(M_*)$. The observations were carried out with the two BOSS spectrographs at Apache Point Observatory \citep{gunn2006, smee2013}. The MaNGA IFUs consist of hexagonal optical fiber bundles containing 19--127 fibres of 2$^{\prime\prime}$ diameter apiece \citep{law2016}, with three-point dithers employed to fully sample the field of view \citep{drory2015,law2015}. The observations were reduced with the MaNGA Data Reduction Pipeline \citep[DRP;][]{yan2016a, law2016, law2021}, prior to analysis with the MaNGA Data Analysis Pipeline \citep[DAP;][]{belfiore2019,westfall2019, law2021}. MaNGA data and analysis products are available on the SDSS science archive server\footnote{\url{https://data.sdss.org/sas/}} and can also be accessed using the Marvin interface \citep{cherinka2019}.

We obtained values of axis ratio ($b/a$), $R_e$ and $M_*$ from the NASA-Sloan-Atlas (NSA) catalog \citep{blanton2011}\footnote{\url{https://data.sdss.org/sas/dr17/sdss/atlas/v1/nsa_v1_0_1.fits}}, where $R_e$ and $b/a$ are elliptical Petrosian values. We restrict to MaNGA galaxies in the Primary+ and Secondary samples, which were observed out to $\sim$ 1.5\,$R_e$ and $\sim$ 2.5\,$R_e$ respectively, and we further restrict to galaxies with observed axis ratios $b/a > 0.6$. We adopt 0.1\,dex uncertainties on $M_*$, representative of typical $M_*$ uncertainties in the MPA-JHU catalog\footnote{Obtained from \url{https://www.sdss4.org/dr17/spectro/galaxy_mpajhu/}}, while assuming 0.05 dex uncertainties on $R_e$ \citep{deugenio2018}; this results in $M_*$ dominating the uncertainties on $\mathit{\Phi_e}$ and $\mathit{\Sigma_e}$.

We obtained emission line maps from the DAP, extracting fluxes and errors for the following features: $\mathrm{H~\alpha}$, $\mathrm{H~\beta}$, [O~\textsc{iii}]$_{5008}$, [N~\textsc{ii}]$_{6585}$, [S~\textsc{ii}]$_{6718}$, [S~\textsc{ii}]$_{6733}$ and [O~\textsc{ii}]$_{3737, 3729}$. These are corrected by the DAP for Milky Way foreground extinction \citep[][Section 2.2.3]{belfiore2019}, by applying the \citet{odonnell1994} reddening law to the maps of \citet{schlegel1998}. We also extracted $\mathrm{H~\alpha}$ equivalent widths ($EW_{H\alpha}$) from the DAP. We used DAP emission line values from non-parametric summed fluxes in all cases. We restricted to spaxels for which $S/N > 3$ for all obtained emission line fluxes. We corrected emission lines for dust using a \citet{fitzpatrick2019} correction curve, assuming an intrinsic Balmer decrement $\mathrm{H~\alpha/H~\beta} = 2.86$.

We selected star-forming spaxels by employing BPT diagnostics \citep{bpt,osterbrock1985,veilleux1987} and by requiring that $EW_{H\alpha} > 14$~\AA, with the latter requirement serving to avoid galaxies with significant diffuse ionised gas (DIG) contamination in their studied regions \citep{lacerda2018,valeasari2019}. For BPT criteria we used the \citet{kauffmann2003} BPT-NII demarcation line and the \citet{kewley2001} BPT-SII demarcation line. 

We estimated gas metallicities for all star-forming spaxels using the RS32 calibrator of \citet{curti2020}. The RS32 indicator is defined as $\mathrm{[S~\textsc{ii}]_{6718,6733} / H~\alpha + [O~\textsc{iii}]_{5008} / H~\beta}$, wherein all ratios have been corrected for dust. We estimate spaxels' $\mathrm{\log(N/O)}$ using the \citet{florido2022} N2O2 calibrator (their Equation 1), wherein N2O2 is defined as $\mathrm{[N~\textsc{ii}]_{6585}/[O~\textsc{ii}]_{3737, 3729}}$ (likewise corrected for dust). The RS32 indicator is double-valued; we therefore assumed all spaxels to be on the upper metallicity branch, while discarding spaxels with RS32 values beyond the \citet{curti2020} fitted range.

\begin{figure*}
\begin{center}
	\includegraphics[trim = 0.5cm 2.4cm 0cm 15cm,scale=0.9,clip]{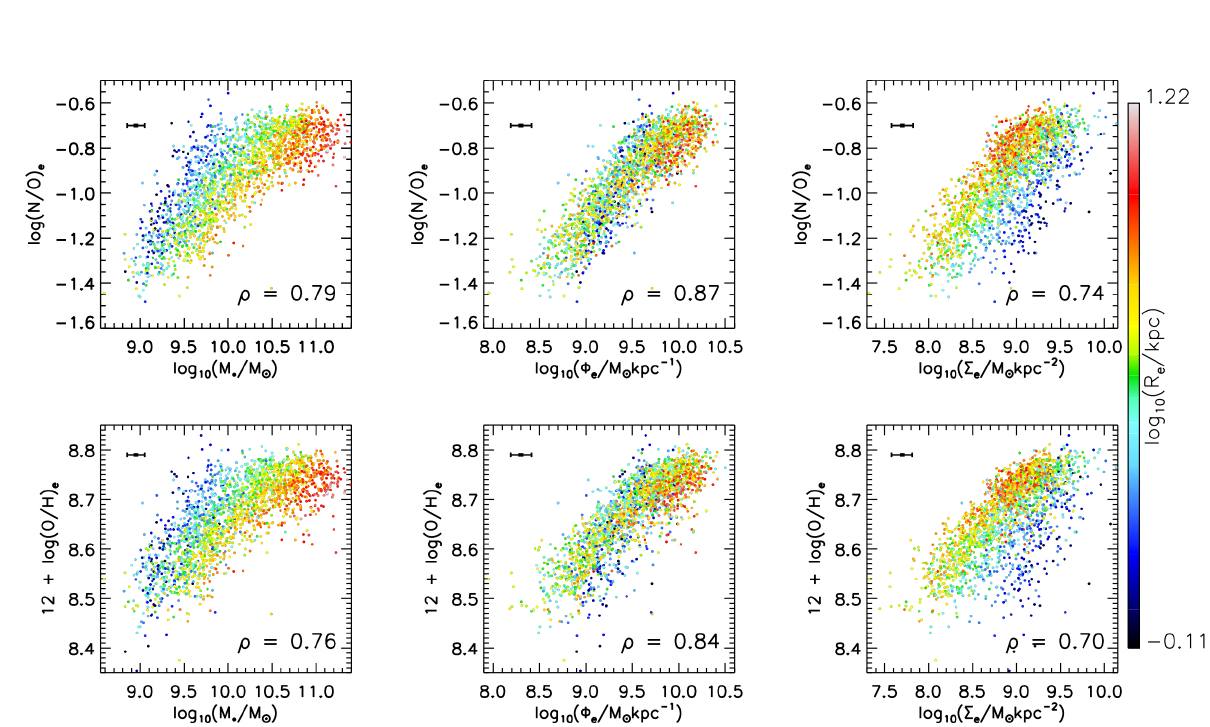} 
	\caption{N/O abundances and gas metallicities, at 1\,$R_e$, as functions of $M_*$ (left panels), $\mathit{\Phi_e}$ (middle panels) and $\mathit{\Sigma_e}$ (right panels). Each panel displays the corresponding Spearman correlation coefficient $\mathit{\rho}$, all with $P \ll 0.01$. We color data points by galaxies' $R_e$, with error bars showing the median uncertainties.}
	\label{mainplot}
	\end{center}
\end{figure*}

We then selected a sample of galaxies for which characteristic chemical abundances can be obtained. Specifically, we aimed to estimate the gas metallicity at 1\,$R_e$, $\mathrm{12 + \log(O/H)_e}$, along with the 1\,$R_e$ N/O abundance $\mathrm{\log(N/O)_e}$. Values at 1\,$R_e$ have been shown to be representative of galaxies' global properties relating to both stars and gas \citep{gonzalezdelgado2014,gonzalezdelgado2015,sanchez2016b}, making them useful as characteristic values. We select our sample by first requiring galaxies to possess at least 40 star-forming spaxels over galactocentric radii 0.5--1.5\,$R_e$ with at least 20 spaxels on either side of the midpoint; this produces a sample of 2070 galaxies. We then grouped a given galaxy's spaxels into a series of elliptical annuli of width 0.1\,$R_e$, covering 0.5--1.5\,$R_e$. We required a minimum of 5 star-forming spaxels per annulus, while also requiring at least 3 annuli \rev{in total} with coverage on both sides of 1\,$R_e$; \rev{we found all 2070 sample galaxies to meet these criteria}. 

We computed galaxies' radial profiles in $\mathrm{\log(N/O)}$ and $\mathrm{12 + \log(O/H)}$ over 0.5--1.5\,$R_e$ by calculating the median $\mathrm{\log(N/O)}$ and $\mathrm{12 + \log(O/H)}$ within each annulus for a given galaxy, before performing an unweighted straight line fit with the \textsc{IDL} \textit{linfit} routine. We used these fits to obtain galaxies' abundances at 1\,$R_e$. Our overall methodology is similar to that employed in the MaNGA pipe3d value-added catalog for gas-phase abundances \citep{sanchez2016,sanchez2016b, sanchez2018, sanchez2022}, though we note that our higher $EW_{H\alpha}$ cut provides somewhat greater resistance to DIG contamination. 

\rev{We obtained errors on our abundances by performing 100 Monte Carlo simulations of each straight-line fit with bootstrapped residuals, with errors reported as the standard deviations from the Monte Carlo fits. Our obtained errors on $\mathrm{12 + \log(O/H)_e}$ and $\mathrm{\log(N/O)_e}$ are small: we find median errors of just 0.002 dex and 0.005 dex respectively, indicating the measured abundances to be highly precise. This occurs due to the large number of star-forming spaxels --- typically hundreds --- that most of our sample galaxies possess, with all ten annuli included in most cases. Considerable systematic uncertainty arises from variations between calibrators \citep[e.g.][]{kewley2008, scudder2021, florido2022}, meaning that \textit{absolute} values of abundances will be much more uncertain than \textit{relative} values of abundances across our sample.} 

For comparison, we also obtained abundances at 1\,$R_e$ by computing radial profiles directly from galaxies' spaxel maps without employing annuli, which were again fit with \textit{linfit}. We found these to produce entirely equivalent results to those from the annulus-based abundances. Thus, we focus on the annulus-based values for the remainder of this article.

 We present the parent and final samples in terms of mass and star-formation rate (SFR) in \autoref{sfms}, which demonstrates our sample selection to implicitly favour galaxies with significant ongoing star-formation. We obtained SFRs from the \textit{log\_sfr\_ssp} column in the Pipe3d summary catalog \citep{sanchez2022}, in which SFRs are derived from spectral fits over the full MaNGA field of view, and we adjusted these SFRs to a Chabrier IMF.

\section{Results}\label{results}

In \autoref{mainplot}, we plot $\mathrm{\log(N/O)_e}$ and $\mathrm{12 + \log(O/H)_e}$, as functions of $M_*$, $\mathit{\Phi_e}$ and $\mathit{\Sigma_e}$. We also give the corresponding Spearman correlation coefficients $\rho$, with \rev{p-value} $P \ll 0.01$ for all quoted values. We find both N/O and metallicity to correlate most tightly with $\mathit{\Phi_e}$, as has been noted for metallicity previously \citep{deugenio2018, sm2024}. We further find that N/O correlates more tightly with all three parameters ($M_*$, $\mathit{\Phi_e}$ and $\mathit{\Sigma_e}$) than does metallicity. Similarly to \citet{deugenio2018}, we color our data points by $R_e$; we see clear residual size dependencies when plotting abundances against $M_*$ or $\mathit{\Sigma_e}$, further emphasising that $\mathit{\Phi_e}$ is the more predictive parameter. 

\begin{figure}
\begin{center}
	\includegraphics[trim = 1.5cm 1cm 1cm 8.5cm,scale=0.8]{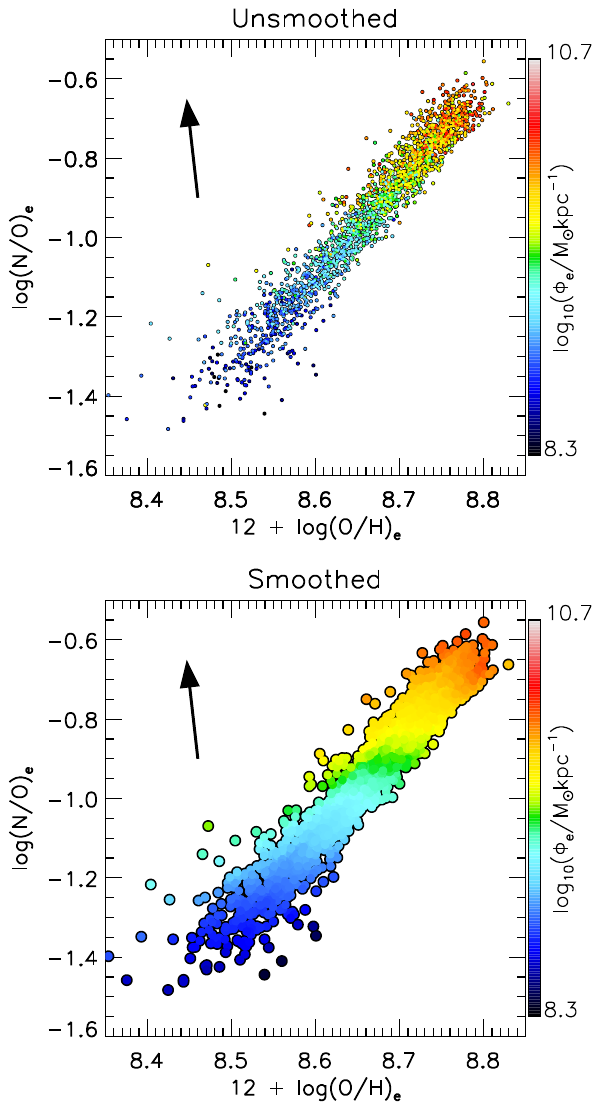} 
	\caption{Potential $\mathit{\Phi_e}$ as a combined function of $\mathrm{\log(N/O)_e}$ and $\mathrm{12 + \log(O/H)_e}$, before (top) and after (bottom) applying LOESS smoothing. The black arrow represents the direction of increasing $\mathit{\Phi_e}$ in N/O--O/H space and is computed via partial correlation coefficients.}
	\label{nooh_pot}
	\end{center}
\end{figure}

In \autoref{nooh_pot}, we consider $\mathit{\Phi_e}$ in chemical space by presenting it as a combined function of $\mathrm{12 + \log(O/H)_e}$ and $\mathrm{\log(N/O)_e}$; the lower panel is a smoothed version of the upper panel, produced via local-weighted regression smoothing \citep[LOESS;][]{cleveland1988} as implemented in \textsc{IDL}\footnote{Available from \url{http://www-astro.physics.ox.ac.uk/~mxc/software/}. We compute smoothed values using the closest 10\% of data points with the \textit{rescale} keyword applied, and errors using the scatter in neighbouring points.}. We find $\mathrm{12 + \log(O/H)_e}$ and $\mathrm{\log(N/O)_e}$ to themselves correlate tightly ($\rho = 0.97$). Thus, we quantify the variation of $\mathit{\Phi_e}$ across chemical space using partial correlation coefficients $\rho_{ij,k}$, which have been applied in multiple recent galaxy studies \citep[e.g,][]{bait2017,bluck2020b,baker2022}; in this notation, the correlation is calculated between parameters $i$ and $j$ with a third parameter $k$ controlled for.

We employ partial correlation coefficients to determine the direction of maximum increase for $\mathit{\Phi_e}$, which we present as an arrow in \autoref{nooh_pot}. We detect a significant correlation between $\mathrm{\log(N/O)_e}$ and $\mathit{\Phi_e}$ with $\mathrm{\log(O/H)_e}$ accounted for, but we detect very little residual trend between $\mathrm{\log(O/H)_e}$ and $\mathit{\Phi_e}$ when $\mathrm{\log(N/O)_e}$ is accounted for. Such a result is also apparent visually from the smoothed $\mathit{\Phi_e}$ values: $\mathit{\Phi_e}$ varies very little with $\mathrm{\log(O/H)_e}$ at a given $\mathrm{\log(N/O)_e}$, in spite of the strong metallicity correlations shown in \autoref{mainplot}.

\subsection{Comparison with fibre-based abundances}

Our use of 1\,$R_e$ annulus abundances is a crucial aspect of this work. Our sample possesses a median $\log(R_e/\mathrm{kpc})$ of 0.61 and hence extends to sizes far above the region of sufficient fibre coverage in \citet{deugenio2018} ($\log(R_e/\mathrm{kpc}) \lesssim 0.5$ at $\log(M_*/M_\odot) > 10$; their figure 5). To further quantify this point, we obtained r-band Petrosian half-light radii ($R_{50}$) from the NSA catalog. We found 889 galaxies (43\%\ of our sample to possess $R_{fib}/R_{50} < 0.5$, where $R_{fib}$ indicates the 3$''$ SDSS fibre radius, which in \citet{deugenio2018} signifies insufficient coverage. Consequently, it can immediately be expected that our results for these galaxies would \textit{not} hold were SDSS fibre-based measurements to be used for this sample.

We investigated fibre measurements by cross-matching our sample with MPA-JHU fibre emission line fluxes, processed in the same manner described for MaNGA spaxel measurements in \autoref{sampledata}. We restricted the test to galaxies satisfying BPT requirements along with satisfying $EW_{H\alpha} > 14$ \AA\ and S/N requirements, which reduces the sample to 754 objects. We found O/H and N/O to correlate more strongly with $M_*$ ($\rho = 0.72, 0.77$ for O/H, N/O) than with $\mathit{\Phi_e}$ ($\rho = 0.66, 0.74$ for O/H, N/O) or $\mathit{\Sigma_e}$ ($\rho = 0.41, 0.47$ for O/H, N/O), which disagrees with our MaNGA analysis. Restricting to the 567 galaxies for which $R_{fib}/R_{50} \geq 0.5$, we found O/H and N/O to correlate more similarly with $M_*$ ($\rho = 0.71, 0.75$ for O/H, N/O) than with $\mathit{\Phi_e}$ ($\rho = 0.68, 0.75$ for O/H, N/O), though we still don't reproduce the findings of \citet{deugenio2018} or of our main analysis. Sample selection differences are a possible reason for the discrepancy with \citet{deugenio2018} in this case. Nonetheless, it is clear that our use of 1\,$R_e$ abundances is an important aspect of our analysis.

\section{Discussion \& Conclusions}\label{disc}

We have demonstrated a tight correlation between the N/O chemical abundance in galaxies and the $\mathit{\Phi_e}$ parameter ($M_*/R_e$). We have further demonstrated that this trend is tighter than previously-reported correlations between $\mathit{\Phi_e}$ and gas metallicity, for which we in fact detect very little connection once N/O is accounted for via partial correlation coefficients.

\rev{Our results, we note, hold over a range of calibrators. In particular, we obtain very similar results if we instead use the \citet{curti2020} O3N2 or R23 calibrators to obtain gaseous metallicities\footnote{These both overlap with N2O2 in terms of employed emission lines, leading us to prefer RS32 for our main analysis.}. Our results are also very similar if we instead use the N2S2 calibrator of \citet{pm2009} (their Equation 22) to obtain N/O abundances. We obtained different results when using metallicities derived from the \citet{curti2020} N2 calibrator or the \citet{Dopita_2016_EmLineDiagnostic} N2S2H$\mathrm{\alpha}$ calibrator: in both cases, we find potential to trend primarily with O/H with little residual N/O dependence when the N2O2 indicator is used. Such a difference is not surprising: the N2 and N2S2H$\mathrm{\alpha}$ calibrators rely heavily on nitrogen emission features and will be insensitive to O/H variations at fixed N/O, making them less appropriate for our analysis than nitrogen-independent indicators such as RS32 or R23 \citep[see also][]{schaefer2020}}\footnote{We note that the form of the N2S2H$\mathrm{\alpha}$ index ($\mathrm{N2S2 - 0.264 N2}$) produces an especially tight correlation with the N2S2 index, making N2S2H$\mathrm{\alpha}$ metallicities particularly ill-suited for considering N/O and O/H together.}.

The $\mathit{\Phi_e}$--metallicity relation has previously been interpreted as reflecting the importance of escape velocity \citep{deugenio2018,sm2024}, with $\mathit{\Phi_e}$ understood as a proxy for a galaxy's gravitational potential. Higher values of $\mathit{\Phi_e}$ would be expected to correspond to higher escape velocities, with higher escape velocities conferring greater resistance to metal-loss via outflows; this in turn is expected to lead to higher measured metallicities \citep[e.g.][]{tremonti2004}.

\citet{baker2023a} however argue for an alternative scenario. They report gas metallicities to correlate more tightly with $M_*$ than with dynamical mass $M_{dyn}$ or with $M_{dyn}/R_e$, where $M_{dyn}$ is derived from the Jeans Anisotropic Mass modelling \citep[JAM;][]{cappellari2008} analysis of \citet{li2018}. \citet{baker2023a} therefore argue for a scenario in which metallicity is driven primarily by the integral of metal-production within a galaxy, with the mass-metallicity relation emerging as a direct consequence.  

The \citet{baker2023a}  scenario effectively treats both $M_*$ and metallicity as time indicators: galaxies grow in mass over time while also enriching over time, thus resulting in more massive galaxies being more metal-rich. It should be noted however that the \citet{baker2023a} scenario cannot \textit{in itself} explain the observed inverse trend between metallicity and size \textit{at a given stellar mass} \citep[e.g.][]{ellison2008,deugenio2018,sm2024}, with the $\mathit{\Phi_e}$ parameter having not been considered in their analysis. Furthermore, this scenario does not directly consider the roles of  gaseous inflows and outflows \citep[e.g.][]{schmidt1963,lilly2013,bb2018, yang2024}, both of which can be expected to lead to scatter in the mass--metallicity relation. These sources of scatter can be understood as follows:

\begin{itemize}
    \item Variations in gaseous inflow rates at a given stellar mass, producing short-term changes in galaxies' gas metallcities.
    \item Variations in gaseous outflow rates at a given stellar mass, driven by variations in escape velocity; this would lead to longer-term differences in galaxies' gas metallicities.
\end{itemize}

Both sources of scatter, we argue, are significantly reduced if one instead considers the $\mathit{\Phi_e}$--N/O relation:

\begin{itemize}
    \item Gaseous inflows can typically be expected to be metal-poor, and so they primarily affect the hydrogen content. Thus, by considering N/O instead of O/H, the scatter in measured relations is reduced.
    \item Escape velocity variations will correspond to variations in gravitational potential, with gravitational potential encoded within the $\mathit{\Phi_e}$ parameter. Thus, by considering $\mathit{\Phi_e}$ instead of $M_*$, the scatter in measured relations is further reduced.
\end{itemize}

\rev{Given the time-delay between oxygen and nitrogen enrichment, star-formation histories (SFHs) are another important consideration \citep[e.g.][]{edmunds1978,molla2006,vincenzo2016,matthee2018}.  $\mathit{\Phi_e}$ can be expected to encode SFH information more effectively than $M_*$, with more compact star-forming galaxies found to be older at a given stellar mass \citep[e.g.][]{li2018,barone2020}. $\mathit{\Phi_e}$ is in fact found to correlate more closely than $\mathit{\Sigma_e}$ or $M_*$ to MaNGA pipe3d \citep{sanchez2022} light-weighted stellar ages at 1\,$R_e$ (Boardman et al. in prep). If higher-$\mathit{\Phi_e}$ galaxies are typically older, with a greater proportion of stars formed at earlier times, then this would confer greater nitrogen production (at fixed metallicity) from previous generations of stars. An N/O--SFH connection could also explain trends between N/O and SFR at fixed $M_*$ \citep{hp2022}, as well as explaining radial abundance profiles of individual galaxies along the N/O--O/H plane \citep{pilyugin2024}.}

Given the above, we argue \rev{the extremely tight $\mathit{\Phi_e}$--N/O relation to be due to several factors}. Firstly, $\mathit{\Phi_e}$ indeed encodes information about galaxy mass and hence about the integral of metal-production, producing a positive $\mathit{\Phi_e}$--N/O correlation. Secondly, $\mathit{\Phi_e}$ encodes information about escape velocities and thus accounts for some of the scatter in the mass--metallicity relation. \rev{Thirdly, variations in $\mathit{\Phi_e}$ \textit{at a given metallicity} will correspond in part to age variations, with older galaxies expected to experience greater nitrogen enrichment at a given metallicity.} Finally, variations in metal-poor inflow rates can be expected to significantly impact O/H abundances on short timescales, with N/O abundances comparatively unaffected (see Section \ref{intro}); this is because metal-poor inflows mostly increase the abundance of hydrogen while having far less impact on heavier elements' relative abundances. Thus, while O/H and N/O both serve as indicators of chemical evolution, N/O is less impacted by short timescale events such as recent inflow, leading to a more robust connection to overall galactic chemical enrichment.

To summarise, we report in this letter a notably tight $\mathit{\Phi_e}$--N/O relation detected in MaNGA galaxies. This relation is tighter than both the mass--metallicity relation and the $\mathit{\Phi_e}$--metallicity relation. The $\mathit{\Phi_e}$--N/O relation also appears more fundamental, in the sense that only a mild $\mathit{\Phi_e}$--metallicity correlation is found once the N/O abundance is accounted for. We argue these results are due to $\mathit{\Phi_e}$ encoding information about both \rev{SFHs} and escape velocities \rev{in addition to $M_*$}, with N/O being relatively \rev{insensitive to} metal-poor inflows.

\section*{Acknowledgements}


NFB and VW acknowledge Science and Technologies Facilities Council (STFC) grant ST/V000861/1. NVA and VW acknowledge the Royal Society and the Newton Fund via the award of a Royal Society--Newton Advanced Fellowship (grant NAF\textbackslash{}R1\textbackslash{}180403). NVA acknowledges support from Conselho Nacional de Desenvolvimento Cient\'{i}fico e Tecnol\'{o}gico (CNPq). Funding for the Sloan Digital Sky Survey IV has been provided by the Alfred P. Sloan Foundation, the U.S. Department of Energy Office of Science, and the Participating Institutions. SDSS-IV acknowledges support and resources from the Center for High-Performance Computing at the University of Utah. The SDSS web site is \url{www.sdss.org}. 

SDSS-IV is managed by the Astrophysical Research Consortium for the Participating Institutions of the SDSS Collaboration including the Brazilian Participation Group, the Carnegie Institution for Science, Carnegie Mellon University, the Chilean Participation Group, the French Participation Group, Harvard-Smithsonian Center for Astrophysics, Instituto de Astrof\'isica de Canarias, The Johns Hopkins University, Kavli Institute for the Physics and Mathematics of the Universe (IPMU) / University of Tokyo, Lawrence Berkeley National Laboratory, Leibniz Institut f\"ur Astrophysik Potsdam (AIP),  Max-Planck-Institut f\"ur Astronomie (MPIA Heidelberg), Max-Planck-Institut f\"ur Astrophysik (MPA Garching), Max-Planck-Institut f\"ur Extraterrestrische Physik (MPE), National Astronomical Observatories of China, New Mexico State University, New York University, University of Notre Dame, Observat\'ario Nacional / MCTI, The Ohio State University, Pennsylvania State University, Shanghai Astronomical Observatory, United Kingdom Participation Group, Universidad Nacional Aut\'onoma de M\'exico, University of Arizona, University of Colorado Boulder, University of Oxford, University of Portsmouth, University of Utah, University of Virginia, University of Washington, University of Wisconsin, Vanderbilt University, and Yale University.

\section*{Data Availability}

All data used here are publicly available.

\bibliographystyle{mnras}
\bibliography{bibliography}





\label{lastpage}
\end{document}